# Nuclear magnetic resonance on a single atom with a local probe

Hester G. Vennema[1,*], Cristina Mier[1,*], Evert W. Stolte[1], Leonard Edens[2], Jinwon Lee[1,†], Sander Otte[1,‡]

[1]Department of Quantum Nanoscience, Kavli Institute of Nanoscience, Delft University of Technology, Delft, The Netherlands
[2]CIC nanoGUNE-BRTA, Donostia-San Sebastián, Spain
[*] These authors contributed equally
[†] jinwon.lee@tudelft.nl
[‡] a.f.otte@tudelft.nl

**Abstract**

The nuclear spin is a prime candidate for quantum information applications due to its weak coupling to the environment and inherently long coherence times. However, this weak coupling also challenges the addressability of the nuclear spin. Here we demonstrate nuclear magnetic resonance (NMR) on a single on-surface atom using a local scanning probe. We employ an electron-nuclear double resonance measurement scheme and resolve nuclear spin transitions of a single $^{47}$Ti isotope with a nuclear spin of $I$ = 5/2. The quadrupole interaction enables to resolve multiple NMR transitions, which are consistent with our eigenenergy calculations. Our experimental results indicate that the nuclear spin can be driven efficiently irrespective of its hybridization with the electron spin, which is required for direct control of the nuclear spin in the long-lifetime regime. This investigation of NMR on a single atom in a platform with atomic-scale control is a valuable development for other platforms deploying nuclear spins for characterization techniques or quantum information technology.

**Introduction**

Nuclear magnetic resonance (NMR) is a powerful characterization technique with applications in physics, chemistry, biology and medicine. Due to the small magnetic moment of nuclear spins and consequential small energy scales, NMR often relies on strong magnetic fields for a large energy splitting and on ensemble averaging. Over the last two decades, enhanced sensitivity of magnetic resonance techniques allowed for sensing individual nuclear spins and nanoscale spatial resolution[1–4]. Moreover, individual nuclear spins are desirable candidates for quantum technology applications[5] since their small magnetic moment contributes to good isolation from their environment, resulting in longer lifetimes and coherence times[6] compared to its electric counterpart. However, this also poses challenges for spin control and therefore viability of nuclear-spin-based quantum technologies. Coherent control of single nuclear spins has been demonstrated on a molecular break junction[7,8], in diamond with nitrogen-vacancy centres[9–15], single atom dopants in silicon[6,16,17] and in crystalline structures via fluorescence detection[18]. Although promising, these platforms lack intervening control of the local atomic environment of the nuclear spin.

Scanning tunnelling microscopy (STM) is a versatile technique which allows for addressing individual atoms on a surface and investigating fundamental interactions of single spins and their environment, such as spin-spin interactions[19–21] as well as magneto-crystalline anisotropies[22,23]. Uniquely, STM offers precise spatial control at subatomic length scales and the ability to build structures through atomic manipulation[24,25] that are tailored to specific Hamiltonians[26,27]. The implementation of electron spin resonance (ESR) in STM[28,29] has improved its energy resolution and offers coherent control of electron spins[21,30,31]. Hence, atomic spins on surfaces probed by STM allow for investigating the fundamental nature of relaxation and decoherence of individual and coupled spins.

Employing the improved energy resolution of ESR-STM, nuclear spins of single isotopes are identified by resolving the hyperfine splitting[32]. Recently, hyperfine-mediated coherent electron-nuclear spin dynamics[33] and single-shot readout of the nuclear spin[34] have been demonstrated. Notably, nuclear spin polarization and addressability have been reported on single Cu and Ti atoms[33,35]. However, the NMR measurements in these studies were carried out under high electron-nuclear spin hybridization, resulting in NMR-like transitions that also involve the electron spin. NMR driving without hybridization with the electron spin has not yet been shown. The ability to independently drive single nuclear spins would signify a key step toward their coherent control.

In this work, we unambiguously demonstrate NMR driving independent of electron-nuclear spin hybridization by employing an electron-nuclear double resonance (ENDOR)[36,37] scheme. We use the ESR signal amplitude for each nuclear spin state ($m_I$) as an NMR spectroscopic signal, since it refers to the time-averaged population of the corresponding nuclear spin state[35]. Leveraging the quadrupole interaction, we resolve different NMR transitions and drive them selectively. The measured NMR energies are corroborated by calculations based on an effective spin Hamiltonian. Investigation of these transitions as function of magnetic field reveals that the driving mechanism potentially originates from an oscillating magnetic field. Our results pave the way to coherently control this long-lived on-surface spin and expand quantum sensing and quantum simulation abilities of STM.

**Results**

**Single-frequency spectroscopy**

We studied titanium (Ti) atoms, carrying an electronic spin $S$ = 1/2, adsorbed onto oxygen binding sites of bilayer MgO grown on Ag(100) (see Methods section). We applied an external magnetic field perpendicular to the surface to lift the electron spin degeneracy and drove ESR by applying a radio frequency (RF) voltage to the tunnel junction (see Fig. 1a for the schematic). Using spin-dependent tunnelling current readout, we detect changes in the magnetoresistance when its frequency is on resonance with the ESR transition. On the studied $^{47}$Ti isotope (nuclear spin $I$ = 5/2), we resolve six different ESR peaks due to hyperfine splitting of the electron spin levels[32] (Fig. 1b, yellow), as opposed to a single peak observed on Ti isotopes without nuclear spin (Fig. 1b, grey). Because of nuclear spin switching during the acquisition time, the relative peak heights of the ESR spectrum directly correspond to the time-averaged nuclear spin occupation probability for each $m_I$ state. The unbalanced population distribution indicates nuclear spin polarization via inelastic electron scattering mediated by the hyperfine interaction[34,38], consistent with previous works[33,34].

The effective spin Hamiltonian for this system can be written as:

$$\hat{H} = \sum_{i=x,y,z} \left( \mu_B g_{e,i}(B_i + B_{tip,i}) \hat{S}_i + \mu_N g_N(B_i) \hat{I}_i + A_i \hat{S}_i \hat{I}_i + Q\eta_i \hat{I}_i^2 \right), \qquad (1)$$

where $\mu_{B(N)}$ is the Bohr (nuclear) magneton and $g_{e(N)}$ is the electron (nuclear) g-factor. $\hat{S}_i$ and $\hat{I}_i$ denote the electron and nuclear spin operators per spatial axis, respectively. $A$ is the anisotropic hyperfine interaction, $Q$ the quadrupole interaction energy, and $\eta$ the quadrupole anisotropy tensor. The first two terms describe the Zeeman energy of the electron and nuclear spin, respectively, where the electron Zeeman splitting has contributions from the external magnetic field ($B$) and the magnetic field produced by the spin-polarized tip ($B_{tip}$). At sufficiently large magnetic field (right hand side of Fig. 1c), the electron Zeeman energy dominates over the hyperfine interaction energy and we can describe the system as a product state ($|m_s, m_I\rangle$) of $m_s$ and $m_I$, the eigenvalues of the $\hat{S}_z$ and $\hat{I}_z$ operators, respectively.

Sweeping the frequency around half of the hyperfine interaction energy ($A_z/2 \sim 65$ MHz) and at low magnetic field reveals features resembling NMR transitions (light purple in Fig. 1d), as previously reported[33]. In this regime, however, $m_I$ and $m_s$ are not good quantum numbers and the eigenstates take the form of hybrid states of $|\downarrow, m_I\rangle$ and $|\uparrow, m_I - 1\rangle$. Moreover, magnetoresistive readout is an electron spin projective measurement, and thus only sensitive to changes in electron spin populations. Therefore, the observed signal in Fig. 1d involves electron spin transitions ($\Delta m_s \neq 0$) and hence, are not pure nuclear spin transitions.

With increasing field, the observed resonance features vanish and are completely absent at 200 mT (dark purple in Fig. 1d). At this field strength, the relevant hybridization term is strongly reduced (see SFig. 1), consistent with the notion that these features are directly related to hybridization. Due to this interdependence, direct NMR driving cannot be independently verified, leaving unanswered whether driving the nuclear spin in a high field regime is possible or not. Moreover, hybridization significantly reduces the lifetime of the nuclear spin[34]. This calls for a measurement scheme in which the driving and readout of the nuclear spin are independent.

**Electron nuclear double resonance**

To surpass this hybridization-dependent readout of the nuclear spin, we used an electron-nuclear double resonance (ENDOR)[36,37] scheme, conceptually similar to electron-electron double resonance demonstrated in STM[39]. As discussed, the relative ESR peak heights (Fig. 1b) give a measure of the nuclear spin populations and can be used as readout signal. As shown in Fig. 1e, we simultaneously operated two RF sources, one in a GHz range (ESR frequency $f_{ESR}$) and the other in a MHz range (NMR frequency $f_{NMR}$), with $V_{ESR}$ and $V_{NMR}$ voltage amplitudes, respectively. Addressing the nuclear spin off resonance (Fig. 1e, grey), we recover the typical distribution, comparable to the yellow curve in Fig. 1b. However, when we chose an $f_{NMR}$ on resonance with one of the nuclear spin transitions (shown in blue), the first two ESR peaks equalize in height, evidencing a change in the nuclear spin occupation probabilities of $m_I = -5/2$ and $m_I = -3/2$. Note that this measurement was performed at a magnetic field of 450 mT, exceeding the field value (200 mT) where in Fig. 1d the signal vanishes. Equalization of the ESR peaks implies that we drive the nuclear spin sufficiently to saturate the two nuclear spin populations.

For NMR spectroscopy, we fixed $f_{ESR}$ while sweeping $f_{NMR}$ (Fig. 2a). In the righthand inset, the used values of $f_{ESR}$ are indicated, showing that we probe the populations of the $m_I = -5/2$ (blue) and $-3/2$ (red) states. In Fig. 2a, we clearly see changes in the populations depending on $f_{NMR}$. The two observed spectroscopic features are attributed to the NMR transitions depending on the electron spin state: $|\downarrow, -5/2\rangle \leftrightarrow |\downarrow, -3/2\rangle$ for ~50 MHz and $|\uparrow, -5/2\rangle \leftrightarrow |\uparrow, -3/2\rangle$ for ~85 MHz. Intuitively, the signal shows out- and inward driving of the nuclear spin population as dips (blue)

and peaks (red), respectively. As a control measurement, we fixed $f_{ESR}$ away from any ESR transition and see no dependence on $f_{NMR}$ (grey). Also, we do not see any spectroscopic feature on an $I = 0$ Ti isotope in this frequency range (see SFig. 2).

A more detailed study as function of both $f_{ESR}$ and $f_{NMR}$ is presented in Fig 2b. We covered the first three ESR peaks, probing the nuclear spin populations of $m_I$ = –5/2, –3/2 and –1/2. Depending on $f_{NMR}$, we see decreasing signal (in red) when we are on resonance with an NMR transition. We observe the NMR resonance frequencies approach $A_z/2 \sim 65$ MHz for increasing $m_I$ until –1/2. Note the minor differences with Fig. 2a in signal strength. We assign this slight variation of nuclear spin polarizations in the equilibrium state to different tip heights, potentially resulting from a non-negligible tunnelling current offset (~10 fA).

In Fig. 2b, the calculated NMR energies from Eq. 1 are indicated with purple markers and bars. Through numerical fitting of our NMR data with MHz precision (~ neV energy resolution), we can accurately fit for $A_z$ = 132.1 ± 0.4 MHz, which is consistent with previous studies[33]. We also find $Q$ = -2.8 ± 0.8 MHz, which is a deviation of ~5% compared to previous reports[32,33] (see Supplementary Note 1 for the fitting procedure). This minor difference could be attributed to a different electric field gradient dependent on the tip shape.

The NMR transitions, marked with roman numerals in the schematic energy diagram in Fig. 2c, match the transitions depicted in Fig. 2b. We resolve at least two pairs of transitions: $|m_s, -5/2\rangle \leftrightarrow |m_s, -3/2\rangle$ and $|m_s, -3/2\rangle \leftrightarrow |m_s, -1/2\rangle$ for both $m_s = \uparrow, \downarrow$. Within each pair the quadrupole contribution, added to the hyperfine energy, lifts the degeneracy resulting in two distinct NMR transitions per $m_s$. Another pair of transitions may be discerned between $f_{ESR}$ = 4.0 and 4.1 GHz corresponding to $|m_s, -1/2\rangle \leftrightarrow |m_s, +1/2\rangle$. For this transition there is no quadrupole contribution and the splitting between the resonance lines is solely the nuclear Zeeman energy. The four remaining transitions (VII, VIII, IX, and X) were not measured because of the low ESR signals of the $m_I$ = +3/2 and +5/2 states due to nuclear spin polarization preferring $m_I$ = –5/2.

**Magnetic field study**

To investigate the influence of hybridization on NMR driving we performed ENDOR measurements as a function of magnetic field strength. Figure 3a shows the ENDOR data at out-of-plane magnetic fields ranging from 200 to 1400 mT, where the hybridization between the electron and nuclear spins is reduced by approximately a factor of 10 (see SFig. 1). For each field value, $f_{ESR}$ was set to the first ESR peak, corresponding to a measurement of $m_I$ = –5/2 population (ESR spectra can be found in the SFig. 4a). The NMR features $|m_s, -5/2\rangle \leftrightarrow |m_s, -3/2\rangle$, found at ~50 MHz and ~85 MHz (labelled I and II), are observed for all fields. This is in contrast with the magnetic field dependence observed in Fig. 1d, where the NMR-like features rapidly disappear beyond $B_z$ = 100 mT. We fit the NMR spectra (see Supplementary Note 1 for the details), and plot their transition energies in Fig. 3b, which reveals the nuclear Zeeman energy depending on the magnetic field. We extract the slope of the found NMR transitions as a function of magnetic field and obtain an experimental nuclear g-factor of 0.37 ± 0.04 (see SFig 5), which differs slightly from the literature value 0.315 measured in bulk for $^{47}Ti$[40].

Notably, for most field values, we also observe the second pair of peaks at ~65 MHz and ~75 MHz, corresponding to the $|m_s, -3/2\rangle \leftrightarrow |m_s, -1/2\rangle$ transitions (labelled III and IV). This means that driving transitions III and IV also affects the population of $m_I$ = –5/2. Considering that the time-

averaged population undergoes continuous spin pumping by spin-polarized current and ESR driving, we understand this population to decrease as a result of reduced spin pumping towards the –5/2 state by draining the $m_I$ = –3/2 to –1/2 state through NMR driving (see schematic in Fig. 3c). For increasing magnetic fields, we see two effects on the amplitudes of the observed NMR transitions. First, the inner features (III, IV) become less pronounced as reduced hybridization decreases the efficiency of spin pumping by lowering the flip-flop transition rate (Fig. 3c and SFig 1). Second, the higher energy transitions (II, IV) are reduced, consistent with assigning these transitions to the $m_s = \uparrow$ states, which are energetically unfavourable for increasing magnetic fields.

This is further corroborated by measuring ENDOR as a function of $V_{ESR}$, as it is sensitive to both nuclear spin and electron spin populations. Figure 4 shows the ENDOR spectra dependent on $V_{ESR}$ while probing the $m_I$ = –5/2 state. We obtain the relative population of $m_s = \downarrow$ and $m_s = \uparrow$ by comparing the normalized amplitude of the NMR features I ($P_\downarrow$) and II ($P_\uparrow$) from a 4-Lorentzian fit (purple). By varying $V_{ESR}$, we change the population ratio of $m_s = \downarrow, \uparrow$[41]. For lower $V_{ESR}$, we have less population in the excited electron state $m_s = \uparrow$ and we see an imbalance between $P_\downarrow$ (blue bar) and $P_\uparrow$ (red bar), favouring the former one. We plot the population ratio $P_\uparrow/P_\downarrow$, approaching 1 for increasing $V_{ESR}$ (Fig. 4 inset). This removes the ambiguity as to which transitions belong to the electron spin ground state.

We note that throughout the studied magnetic field range, the ENDOR measurements shown in Fig. 3a vary in signal strength. Due to experimental limitations of the RF transmission line, we were unable to send constant amplitude for all ESR frequencies, especially for fields above 500 mT (see Supplementary Note 2 and SFig. 6). As shown in Fig.4, lower $V_{ESR}$ reduces the ENDOR signal. From this, and from the observation of a non-monotonic decrease of the signal with magnetic field, we determine the reduced RF amplitude being the dominant factor in the decrease of the signal-to-noise ratio (SNR) for higher fields.

**Discussion**

From the observed NMR spectra and their dependence on external magnetic field, we hypothesize possible driving mechanisms of the nuclear spin in this platform. There are three terms in the Hamiltonian (Eq. 1), modulation of which can cause $\Delta m_I = \pm 1$ transitions: the hyperfine interaction, the quadrupole interaction, and the Zeeman interaction. Modulation of the hyperfine interaction would drive flip-flop transitions ($|m_s, m_I\rangle \leftrightarrow |m_s \pm 1, m_I \mp 1\rangle$) involving also the electron spin resulting in transition energies in the GHz range for the magnetic fields used in this work. However, the measured NMR remain in MHz frequency range throughout the studied fields. This allows us to rule out modulation of the hyperfine interaction as the origin for these transitions.

Modulation of the quadrupole interaction, induced by the RF field changing the electric field gradient, allows transitions matching the observed NMR frequencies. But selection rules prohibit to drive the $|m_s, -1/2\rangle \leftrightarrow |m_s, +1/2\rangle$ transition through a quadrupole modulation[16]. We do, however, resolve some response at the expected frequencies of these transitions in Fig. 2b. Moreover, quadrupole modulation would additionally drive $\Delta m_I = \pm 2$, but those transitions are not clearly visible in our measurements (see Supplementary Note 3 and SFig. 8). Therefore, we conclude that a quadrupole modulation cannot drive the nuclear spin solely in our experiments.

Taking this into account, we propose that driving the nuclear spin is dominantly caused by an effective oscillating magnetic field induced via an RF electric field. This could arise in a similar fashion as driving the electron spin in ESR-STM, where a piezo electric displacement of the atom[28,42], modulation of the tunnelling barrier[43,44] or time-dependent spin-torque[45] have been proposed as possible driving mechanisms. Such an external driving mechanism, e.g. not dependent on system specific constants $A$ or $Q$, would imply that nuclear spin driving with an STM can be applied to other nuclear spin systems.

**Conclusion**

In conclusion, we demonstrate nuclear magnetic resonance on a single on-surface atom by performing ENDOR experiments with an STM. This technique separates driving and readout of the nuclear spin, and we resolve the NMR spectrum of a single $^{47}$Ti atom. The presence of quadrupole interaction lifts the degeneracy of the NMR transitions and allows for individual addressability of nuclear spin states. An improved method of state initialization through coherent driving instead of spin scattering could open up the possibility to control the nuclear spin across the full Hilbert space[17,46]. Our results suggest that hybridization of the nuclear spin with the electron spin is not necessary to drive NMR on the single atom. Based on our observations, we expect that the driving of the nuclear spin occurs mainly through an oscillating magnetic field.

This work is a key step towards coherent control of a nuclear spin system with STM. We have shown the ability to drive the nuclear spin at high magnetic fields, where the relaxation time is in order of seconds for this system[34], potentially resulting in coherence times in the same order of magnitude. We anticipate the measurement of the coherence time through pulsed experiments as well as proof of coherent control of the nuclear spin. This development expands the abilities of STM for quantum sensing, simulations, and quantum operations. Additionally, the combination of ENDOR and scanning probe microscopy shown in this work could be used to study decoherence sources of nuclear spins in a controlled environment.

**Methods**

MgO/Ag sample including Ti and Fe surface atoms were prepared with the same method described in previous works[33,34]. Spin-polarization of the tungsten tip was achieved by first indenting the tip in clean Ag until sharp and subsequently picking up ~10-20 Fe atoms. Each ESR-active tip was tested for sufficient nuclear spin polarization dependent on setpoint bias or current at high external field (1400 mT). All data presented in the main text were taken with the same microtip, except for Fig. 1d, which was taken with a different microtip.

The experiments were performed in a commercial Unisoku USM-1300 STM. All measurements were carried out at a sample temperature of 0.4 K. The pre-amplifier for the tunnelling current used for all measurements is an NF corp. SA608F2, with its internal analog 300 Hz low-pass filter active, powered by an NF corp. LP5393 as a low noise DC power supply. We used two Rhode & Schwarz SMA100B signal generators for radio-frequency voltages, which we combined using a splitter/combiner from Minicircuit (ZFRSC-183-S+). Both RF voltages were together combined with the DC bias delivered by a Nanonis V5 Digital-to-Analog converter using a bias tee Tektronix PSPL5542. For all measurements we use a Stanford Research SR830 lock-in amplifier for detection. We chop the RF signal at 271 Hz. In single-tone mode (Figs. 1b and 1d), we sync the used RF generator with the lock-in amplifier. In ENDOR mode, we always chop the GHz signal while the MHz signal is continuous. All RF voltages are expressed in zero-to-peak amplitude. See SFig. 9 for the set-up and different measurement configurations.

**Data availability**

The raw data generated in this study as well as the analysis and simulation code have been deposited in a Zenodo database under the identifier [URL].

**Acknowledgements**

This work was supported by the European Research Council (ERC Advanced Grant No. 101095574 "HYPSTER") and the research program "Materials for the Quantum Age" (QuMat). This program (registration number 024.005.006) is part of the Gravitation program financed by the Dutch Ministry of Education, Culture and Science (OCW).

We acknowledge Irene Fernández de Fuentes, Lukas M. Veldman, Susanne Baumann and Allard Katan for fruitful scientific discussion.

**Author contributions**

H.G.V., C.M., E.W.S., J.L., and S.O. conceived the experiment. H.G.V. and C.M. performed the experiment. H.G.V., C.M., and L.E. analysed the experimental data. H.G.V., C.M., E.W.S., L.E., J.L., and S.O. discussed and interpreted the results. H.G.V., C.M., J.L., and S.O. wrote the manuscript, with input from all authors. S.O. supervised the project.

**Competing interests**

The authors declare no competing interests.

**Materials & Correspondence**

Correspondence and requests for materials should be addressed to Sander Otte or Jinwon Lee.

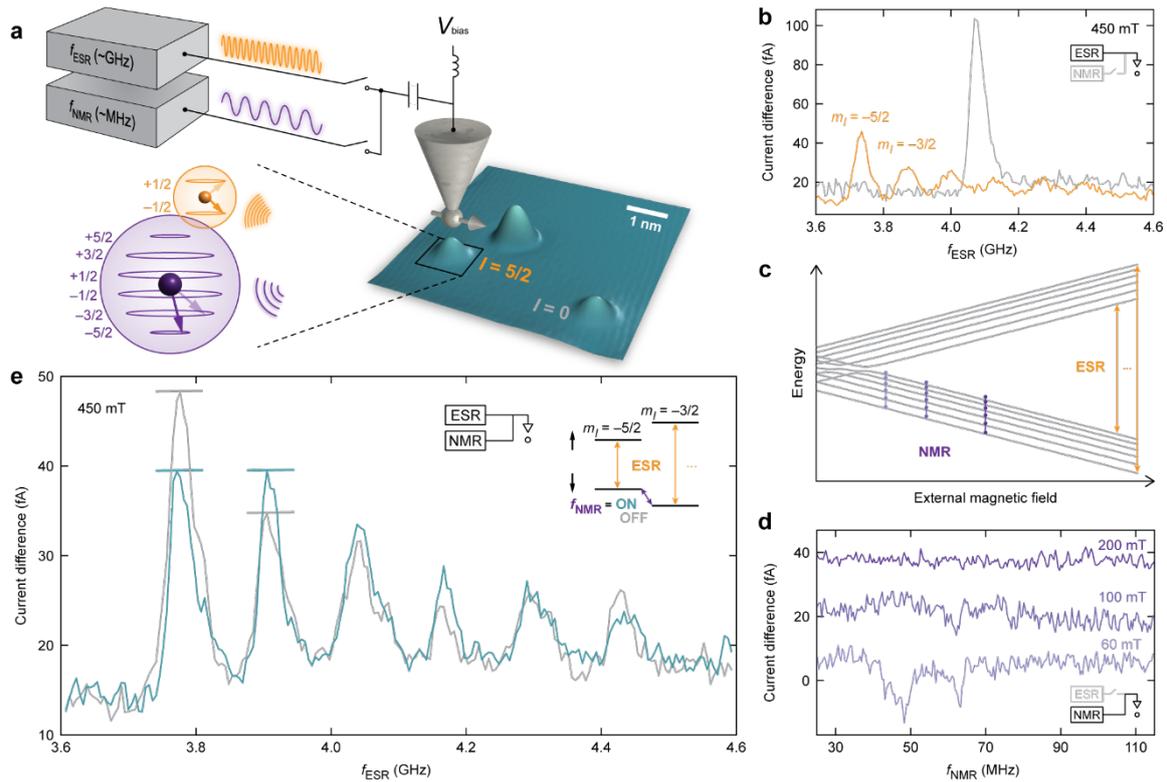

**Fig. 1 | ENDOR measurement scheme. a**, Schematic of the experiment. Two RF sources are combined with the DC bias through a bias tee and connected to the STM tip. With a spin-polarized tip we studied single $^{47}$Ti atoms on MgO/Ag(100) surface. The topography shows the studied atom with nuclear spin ($I = 5/2$) and reference Ti atom without nuclear spin ($I = 0$). We performed experiments in single-tone mode ((b), (d)) or ENDOR mode (e). **b**, Single-tone ESR spectroscopy of two Ti atoms with $I = 0$ (grey) and $I = 5/2$ (orange). $I_{set}$ = 4 pA, $V_{bias}$ = 80 mV, $V_{ESR}$ = 20 mV, $B_z$ = 450 mT. **c**, Energy diagram of the coupled nuclear spin ($I = 5/2$) and electron spin ($S = 1/2$) system as a function of external magnetic field, visualizing relative energy scales of the electron and nuclear spin transitions. The yellow lines indicate ESR transitions observed on the yellow spectra in (b). The purple lines indicate the fields at which the data shown in (d) was measured. **d**, Single-tone NMR spectroscopy at different external magnetic fields. Spectra are vertically shifted for clarity. In the dark purple trace (200 mT) no signal is measured, while in lighter purple (100 mT and 60 mT) NMR-type transitions are observed. For 60 mT: $I_{set}$ = 3 pA, $V_{bias}$ = 75 mV; for 100 mT: $I_{set}$ = 3 pA, $V_{bias}$ = 70 mV; for 200 mT: $I_{set}$ = 2.5 pA, $V_{bias}$ = 90 mV. For all fields: $V_{NMR}$ = 20 mV. **e**, ENDOR spectra with $f_{NMR}$ on (49.3 MHz, blue curve) and off (30.2 MHz, grey curve) resonance with a nuclear spin transition. Horizontal bars are guides to the eye, showing the populations of the first two ESR peaks equalize when $f_{NMR}$ is on resonance. Inset: diagram showing the relevant driving between energy levels: $f_{ESR}$ sweeps and drives several ESR transitions consecutively and $f_{NMR}$ is set to drive $|\downarrow, -5/2\rangle \leftrightarrow |\downarrow, -3/2\rangle$. $I_{set}$ = 4 pA, $V_{bias}$ = 80 mV, $V_{ESR, NMR}$ = 20 mV, $B_z$ = 450 mT.

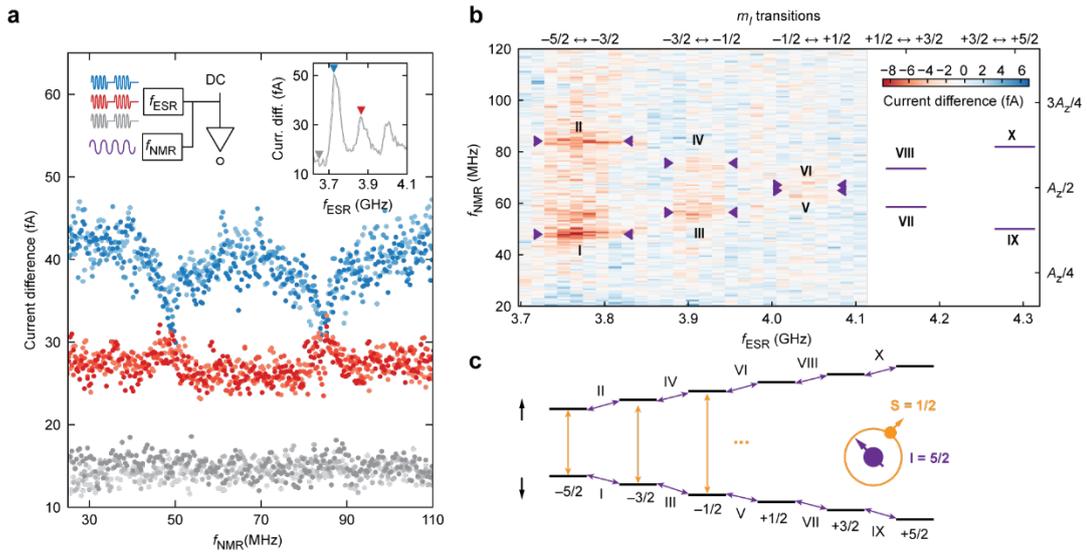

**Fig. 2 | NMR spectroscopy on $^{47}$Ti using ENDOR. a**, ENDOR spectroscopy for three different $f_{ESR}$ frequencies. The colours of the spectra match the indicators in the right inset. The left inset shows the ENDOR measurement scheme: $f_{ESR}$ was fixed to probe different nuclear spin populations and we swept $f_{NMR}$. The ESR signal was chopped for lock-in detection. For each $f_{ESR}$, three sweeps are shown to give an impression of the noise floor. $I_{set}$ = 4 pA, $V_{bias}$ = 80 mV, $V_{ESR, NMR}$ = 20 mV. **b**, ENDOR measurements where we swept $f_{ESR}$ and $f_{NMR}$ to obtain a 2D map of the nuclear spin transitions. Background subtraction has been applied (minus average of each column, see SFig. 3 showing the raw data). The NMR transitions show a decrease in signal and are visible in red. The horizontal axis shows the fixed ESR frequencies, and the vertical axis shows the swept NMR frequencies. $I_{set}$ = 4 pA, $V_{bias}$ = 80 mV, $V_{ESR, NMR}$ = 20 mV, $B_z$ = 450 mT. The calculated NMR frequencies are overlaid with purple arrows and horizontal lines for $f_{ESR}$ > 4.1 GHz. The NMR frequencies are predominantly determined by $A_z/2$, indicated by markers on the right vertical axis, while the quadrupole interaction lifts the degeneracy of the NMR transitions. **c**, Eigenstate diagram of the coupled spin system, ordered in energy. Small black arrows indicate the electron spin state. The quantum number $m_I$ is indicated in type. Double arrows indicated driving. ESR transitions in yellow, NMR transitions in purple.

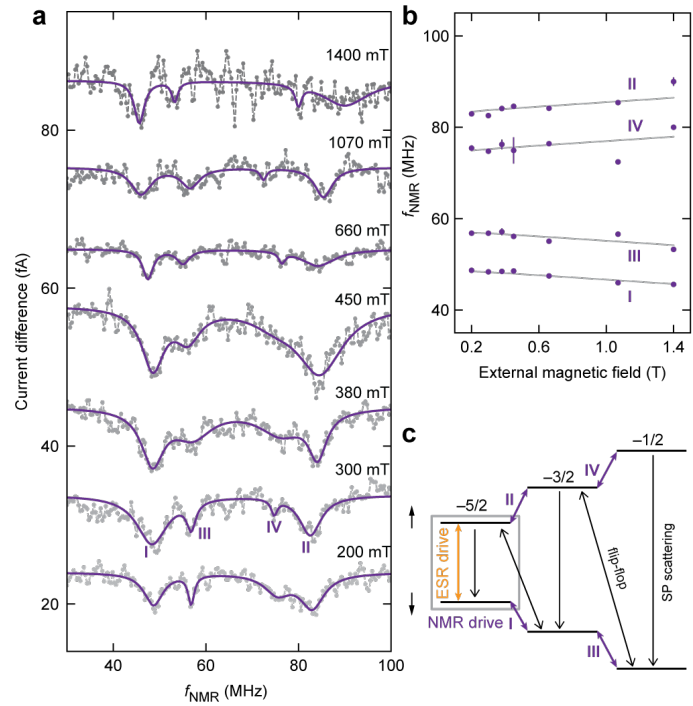

**Fig. 3 | ENDOR measurements external magnetic field study. a**, ENDOR measurements taken at different magnetic fields, all measured with $f_{ESR}$ fixed to the first ESR peak ($m_I = -5/2$), see SFig. 4 for ESR spectra. Each curve is averaged over three sweeps and offset arbitrarily for visibility. The purple solid lines are fits using a four-Lorentzian function (see Suppl. Note 1). For all spectra: $I_{set}$ = 4 pA, $V_{bias}$ = 80 mV, $V_{NMR}$ = 20 mV. For 200-450 mT: $V_{ESR}$ = 20 mV, for 660-1070 mT: $V_{ESR}$ = 8-10 mV, for 1400 mT: $V_{ESR}$ = 12 mV (see Suppl. Note 2). **b**, Fitted NMR peak positions for the studied magnetic field range. Alongside, the NMR energies for transitions I-IV are calculated from Eq (1) and shown in grey dashed lines showing the nuclear Zeeman splitting. **c**, Schematic of the relevant NMR transitions observed in (a). The grey box indicates the $m_I = -5/2$ subspace for which we measure the time-averaged nuclear spin population. Under the influence of continuous-wave ESR driving (yellow arrow) and constant spin pumping through spin-polarized (SP) scattering and spontaneous flip-flop transitions (combined process of black arrows) the population is at a steady-state equilibrium.

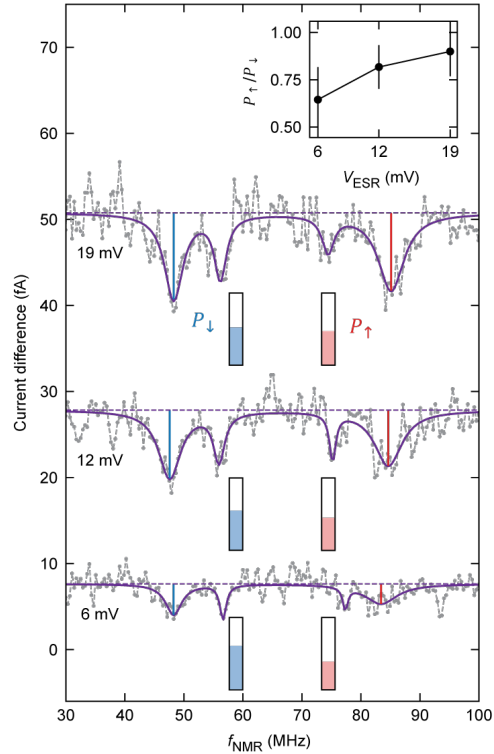

**Fig. 4 | ENDOR as a function of ESR driving power.** ENDOR measurements obtained with different $V_{ESR}$ values with $f_{ESR}$ fixed to the first ESR peak at 3.8 GHz ($m_I = -5/2$). The purple solid lines are fittings with a four-Lorentzian function. The blue and red vertical lines indicate the signal intensity of the $|\downarrow, -5/2\rangle \leftrightarrow |\downarrow, -3/2\rangle$ (I) and $|\uparrow, -5/2\rangle \leftrightarrow |\uparrow, -3/2\rangle$ (II) NMR transitions. The blue and red colour bars depict the normalized signal of I and II ($P_\downarrow$ and $P_\uparrow$), showing an increasing imbalance for lower $V_{ESR}$. *Inset*: $P_\uparrow/P_\downarrow$ as a function of $V_{ESR}$. The reduced ratio with decreasing $V_{ESR}$ indicates a higher occupation of the $m_s = \downarrow$ state. $I_{set}$ = 4 pA, $V_{bias}$ = 80 mV, $V_{NMR}$ = 20 mV, $V_{ESR}$ = 6, 12, 19 mV (bottom to top), $B_z$ = 450 mT.

# Supplementary Information for
# Nuclear magnetic resonance on a single atom with a local probe

Hester G. Vennema[1,*], Cristina Mier[1,*], Evert W. Stolte[1], Leonard Edens[2], Jinwon Lee[1,†], Sander Otte[1,‡]

[1]Department of Quantum Nanoscience, Kavli Institute of Nanoscience, Delft University of Technology, Delft, The Netherlands
[2]CIC nanoGUNE-BRTA, Donostia-San Sebastián, Spain
* These authors contributed equally
† jinwon.lee@tudelft.nl
‡ a.f.otte@tudelft.nl

**Supplementary Note 1 | Spin Hamiltonian analysis.**

We fit the experimental ESR and NMR resonant frequencies by comparing them with numerically calculated values from the effective spin Hamiltonian (Eq. 1). By using values for $\boldsymbol{g}_e$, $\boldsymbol{A}$ and $Q$ reported in literature[1,2] for the studied system (Ti atom adsorbed onto oxygen binding site of MgO and with nuclear spin $I$ = 5/2) in our numerical calculation, we found a slight discrepancy with the experimentally obtained ESR and NMR spectra. In the following, we present a recursive fitting procedure to find appropriate values for the mentioned parameters. Since the external field is oriented out-of-plane, from the fitting, we can extract the z-components of the following parameters: $g_{e,z}$, $A_z$ and $Q$, as well as the magnetic field arising from the spin polarized tip, $\boldsymbol{B}_{tip}$. Here, $g_{e,z}$ is the slope of the Zeeman splitting, $A_z$ is the hyperfine coupling which defines the splitting between ESR resonant peaks in first order, and the quadrupole interaction $Q$ can be seen as a correction to the hyperfine splitting dependent on $m_I$.

First, we analyse the ESR spectra to obtain $g_{e,z}$ and $\boldsymbol{B}_{tip}$. We extract the measured ESR frequencies by fitting the data with a function consisting of six Fano resonances from spectra obtained at fields ranging from 200 to 1400 mT. We assume that the ESR peaks are equally spaced around a centre frequency $f_0$ (we neglect quadrupole interaction at this stage) and that the nuclear spin population (peak intensity) follows a Boltzmann distribution[3], see SFig. 4a. We fit the centre frequency $f_0$ for each spectrum, with a linear function of $B_z$ to extract $g_{e,z}$ and $B_{\text{tip},z}$, which determines the Zeeman splitting at zero external field:

$$f_0(B_z) = \mu_B g_{e,z}(B_z + B_{\text{tip},z}) \quad\quad\quad (S1)$$

As shown in SFig. 4b, we find $g_{e,z}$ = 0.56 ± 0.02 and $B_{\text{tip},z}$ = 65.64 ± 0.02 mT. The obtained z-component of the $g$-factor differs from previously reported measurements[2] by about 5%. $\boldsymbol{B}_{tip}$ can, in principle, be oriented anywhere in the 3D space, described by the field magnitude $B_{\text{tip}}$, $\phi$ (the angle with respect to z-axis), and $\theta$ (the azimuthal angle). We can take $\theta$ = 0 without loss of generality. We also neglect the effect of the tip field on the nuclear Zeeman splitting, since most of this effective field is caused by exchange interaction and hence cannot couple directly with the nuclear spin[4].

To extract $B_{\text{tip}}$ and $\phi$, we fit the expected ESR frequencies obtained by solving the Hamiltonian (Eq. 1 without quadrupole interaction) to the resonances extracted from the Fano fitting. To solve the Hamiltonian, we input the obtained $g_{e,z}$ and $A_z$ = 130 MHz which is taken from literature[1]. $B_{\text{tip}}$ and $\phi$ are fitting parameters, we use $B_{\text{tip}}$ = $B_{\text{tip},z}$ as initial guess and $\phi$ = [0.1, 20˚] as bounds since we assume a primarily out-of-plane tip field. We find $B_{\text{tip}}^* = 67.3 \pm 2.1$ mT and $\phi^* = 5 \pm 2°$, which are similar values to ones found in previous studies[3]. The '*' superscript indicates that these quantities will be further optimised.

In a second step, we compare the measured NMR spectra (Fig. 3a) with transition frequencies calculated with the full Hamiltonian (Eq. 1) to obtain accurate fittings for $A_z$ and $Q$. We consistently observe four NMR features at different external field values, so we use a four-Lorentzian function to fit each spectrum and extract the experimentally measured NMR frequencies, as is shown in Fig. 3a in purple solid lines.

We define the quadrupole contribution as[1,5]:

$$Q\boldsymbol{\eta} = \kappa \frac{1}{2I(2I-1)} \begin{pmatrix} -\frac{1}{2}(1-\eta) & 0 & 0 \\ 0 & -\frac{1}{2}(1+\eta) & 0 \\ 0 & 0 & 1 \end{pmatrix} \quad (S2)$$

Here, $\kappa = \frac{e^2 qQ'}{h}$, where $q$ is the electric field gradient and $Q'$ is the nuclear electric quadrupole moment. For $^{47}$Ti we have $I = 5/2$ and we assume $\eta = 0$ since here the oxygen binding site is 4-fold symmetric and therefore $Q\boldsymbol{\eta}$ is isotropic in x and y.

We perform linear least squares fitting procedure where $A_z$ and $\kappa$ are fitting parameters, and $B_{\text{tip}}^*$, $\phi^*$ and $g_{e,z}$ are fixed. We find good agreement with previous studies[1,3] for $A_z^* = 132.1 \pm 0.4$ MHz and $\kappa^* = -56.7 \pm 0.8$ MHz, resulting in $Q = -2.8 \pm 0.8$ MHz for $I = 5/2$.

We check convergence of our fitting procedure by plugging $A_z^*$ into the Hamiltonian instead of the previously used literature value, and repeat the fitting of $B_{\text{tip}}$ and $\phi$. We find $B_{\text{tip}} = 67.9 \pm 2.1$ mT and $\phi = 5 \pm 2°$, in good agreement with previously found values $B_{\text{tip}}^*$ and $\phi^*$. Afterwards, we also repeat the fitting for $A_z$ and $\kappa$. The obtained values, reported in the main text, differ in less than 0.002% compared to $A_z^*$ and $\kappa^*$, therefore showing converge in our recursive fitting after two steps.

Moreover, we can fit the NMR frequencies from the Lorentzian fitting with a linear slope to extract the nuclear g-factor, $g_N$, see SFig. 5. We find $g_N = 0.37 \pm 0.04$, which is close to the literature value[6] (measured in bulk) for $^{47}$Ti: $g_N = 0.315$. Given the discrepancy with literature, we used the literature value in the fitting procedure described above.

**Supplementary Note 2 | RF transmission**

For the full frequency range (MHz and GHz), we have made appropriate transfer functions for our radio-frequency transmission line and set-up depicted in SFig 9. Due to frequency-dependent attenuation in the transmission line, the RF generator needs to send more power to have the same voltage in the junction ($V_{\text{ESR}} = 20$ mV) for $f_{\text{ESR}} > 5$ GHz, as for $f_{\text{ESR}} < 5$ GHz. We were limited to sending +10 dBm power from the source for our experiments due to increased noise levels and heating effects. Therefore, we could send less $V_{\text{ESR}}$ amplitude to the junction for certain frequency ranges (see SFig. 6). This led to reduced ENDOR signal amplitude for measurements where $B_z > 660$ mT, for which the ESR frequency send were above 5 GHz. This is consistent with a study of the influence of $V_{\text{ESR}}$ amplitude for ESR frequencies on ENDOR data at a constant field (450 mT), see SFig. 7.

**Supplementary Note 3 | Absence higher energy features**

A driving mechanism via quadrupole modulation would also support transitions of $\Delta m_I = \pm 2$, because of the $\sum_{i=x,y,z} \hat{I}_i^2$ operator[7,8]. Therefore, we increase the range of $f_{\text{NMR}}$ to study whether we see features at higher frequencies, the resulting spectrum is shown in SFig. 8. From our system Hamiltonian (Eq. 1) we calculated the expected frequencies by just adding subsequent single NMR transition energies: I+III, III+V, ... and II+IV, IV+VI etc. The frequencies of single NMR transitions as well as double NMR transitions are plotted as vertical lines in SFig. S8. Here, we can observe the single NMR transitions I and III clearly, but features at higher frequencies are absent. This, in combination with the slight signal $|m_s, -1/2\rangle \leftrightarrow |m_s, +1/2\rangle$ shown in Fig. 2b, makes a quadrupole modulated driving mechanism on its own not likely.

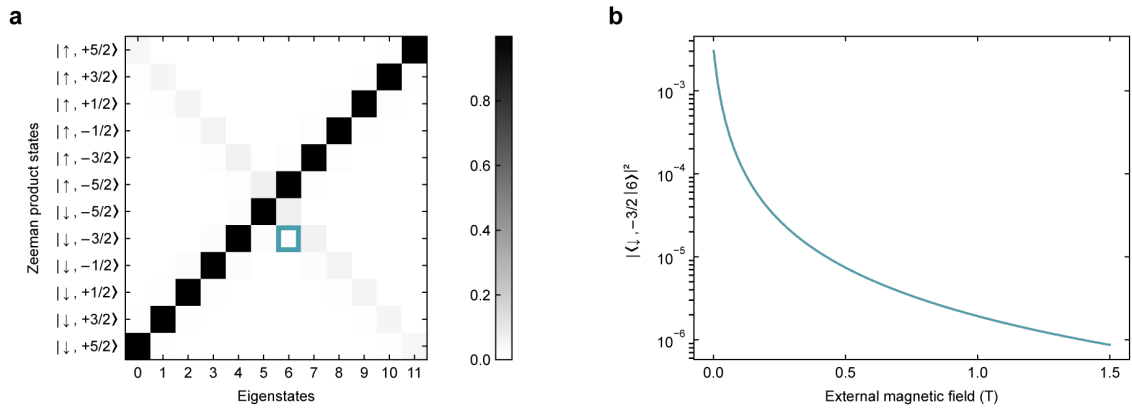

**Supplementary Fig. 1 | a,** Projection of the eigenstates (0-11, ordered in eigenenergy from low to high) of the coupled spin system on the Zeeman product state basis at a fixed external magnetic field of 100 mT. In blue, the element is depicted which resembles the relevant flip-flop coefficient. **b,** Calculation of the off-diagonal matrix element labelled blue in (a), as a function of external magnetic field.

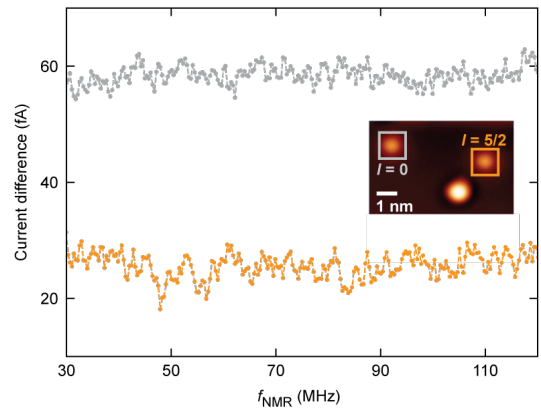

**Supplementary Fig. 2 | ENDOR measurement comparison with *I* = 0 isotope.** ENDOR spectra obtained by fixing $f_{ESR}$ to the single ESR resonant peak (grey) and to the first hyperfine peak (yellow) (see also Fig. 1b). For the grey spectrum (*I* = 0 isotope), we see no spectral feature, while we do see dips around 50 MHz for the orange spectrum (*I* = 5/2 isotope). Inset: topography of the studied atoms (same as in Fig. 1a but rotated 180 degrees).

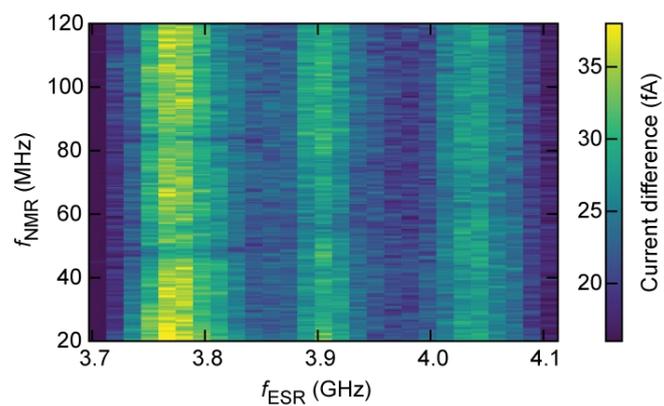

**Supplementary Fig. 3 | ENDOR 2D map raw data.** Same data as shown in Fig. 2b. The three vertical high intensity bands correspond to the first three ESR peaks and relative nuclear spin populations of $m_I$ = −5/2, −3/2 and −1/2.

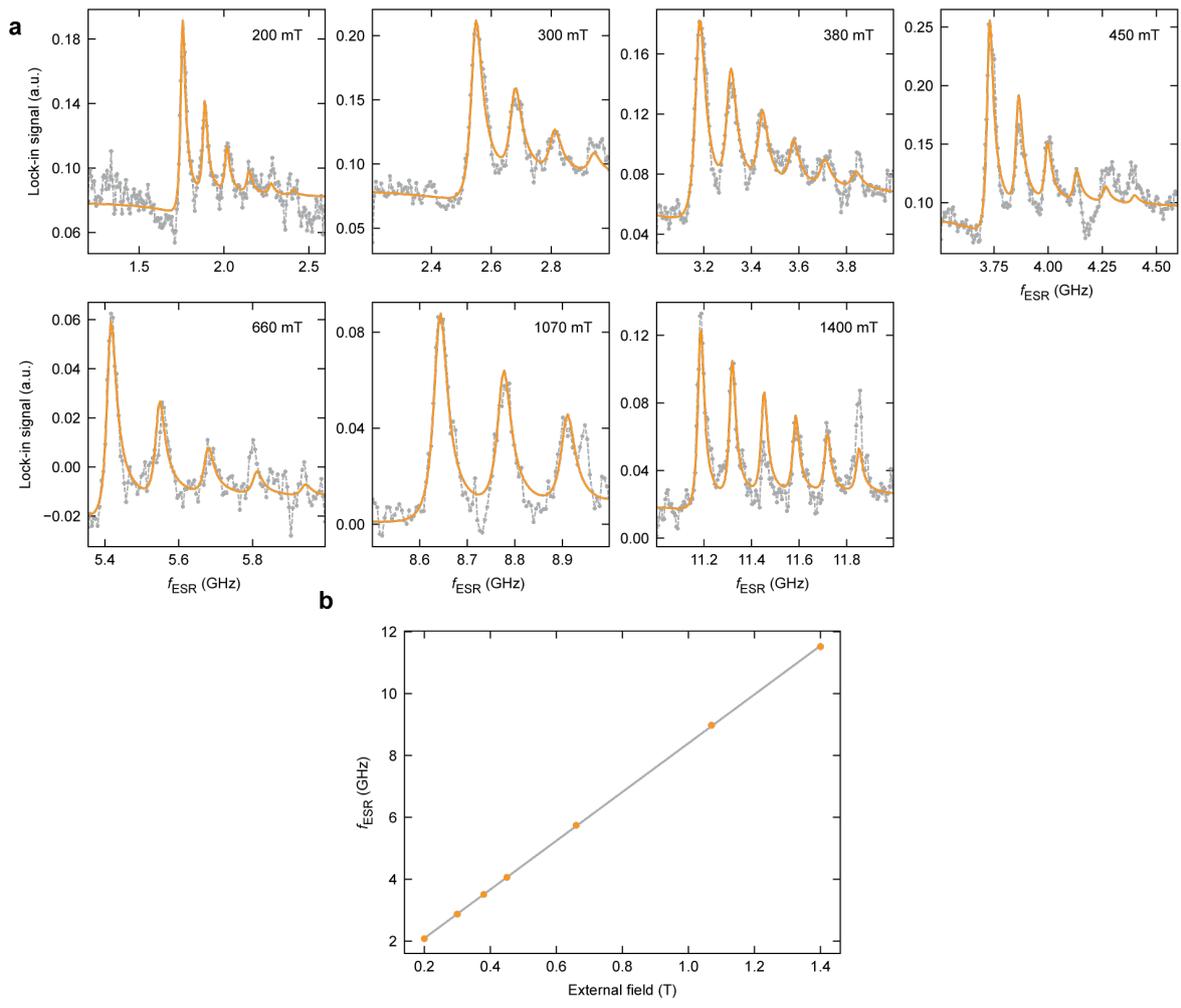

**Supplementary Fig. 4 | Fitting of the ESR spectra. a,** Each spectrum is fitted with 6 Fano functions to extract the frequency of each resonant peak. $I_{set}$ = 4 pA, $V_{bias}$ = 80 mV; $V_{RF}$ varies per field, see SFig. 6b. **b,** Centre frequencies of each hyperfine spectrum per magnetic field are fitted with a linear fit to extract $g_{e,z}$.

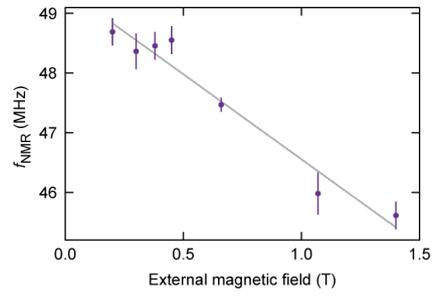

**Supplementary Fig. 5 | Nuclear Zeeman splitting revealed by ENDOR.** Frequencies of the NMR transition $|\downarrow, -5/2\rangle \leftrightarrow |\downarrow, -3/2\rangle$ plotted versus magnetic field shown in purple including error bars of the Lorentzian fitting as shown in Fig. 3a. These points are fitted with a linear function (grey) from which we extract $g_N$ = 0.37 ± 0.04.

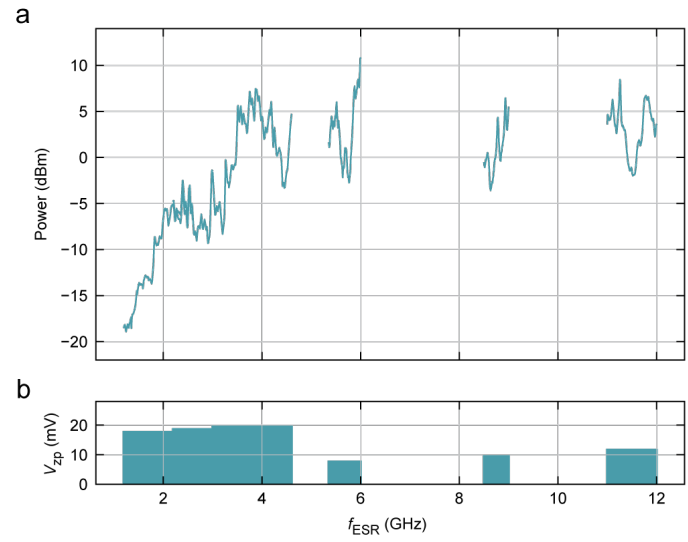

**Supplementary Fig. 6 | RF transmission for ESR frequencies. a,** Sent power values for radio-frequency voltages in the used frequency range for ESR measurements shown in Fig. S3 a. **b,** Corresponding voltage amplitude (zero-to-peak) in the tip-sample junction calculated from transfer functions.

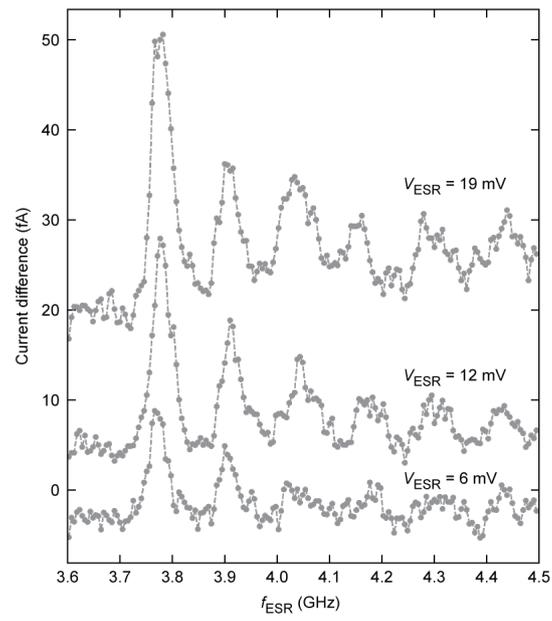

**Supplementary Fig. 7 | ESR measurement as a function of ESR power.** ESR spectra on a $^{47}$Ti isotope measured at $V_{ESR}$ = 6, 12 and 19 mV. $I_{set}$ = 4 pA, $V_{bias}$ = 80 mV, $B_z$ = 450 mT.

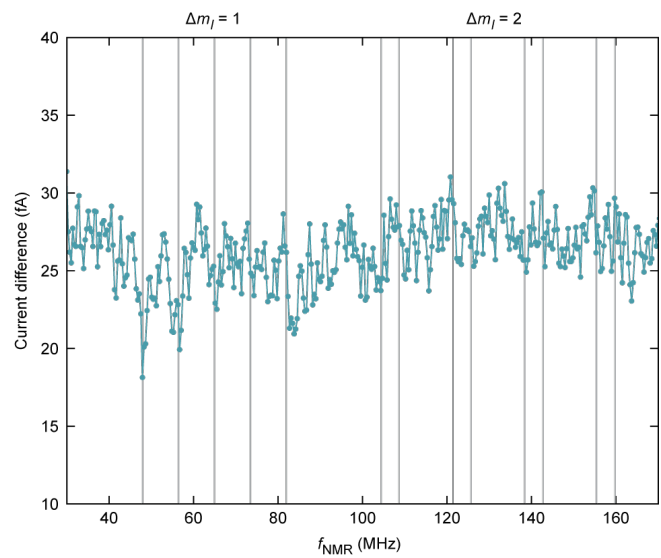

**Supplementary Fig. 8 | Absence of higher energy transitions.** ENDOR measurement for higher frequency where $f_{ESR}$ is fixed to the first hyperfine peak. In grey lines the calculated NMR energies are plotted for single ($\Delta m_I = \pm 1$) and double ($\Delta m_I = \pm 2$) NMR transitions. The calculations show agreement with some $\Delta m_I = \pm 1$ features but no clear signal could be observed for the calculated $\Delta m_I = \pm 2$ transitions. $B_z$ = 450 mT, $I_{set}$ = 4 pA, $V_{bias}$ = 80 mV, $V_{ESR}$ = 19 mV, $V_{NMR}$ = 20 mV.

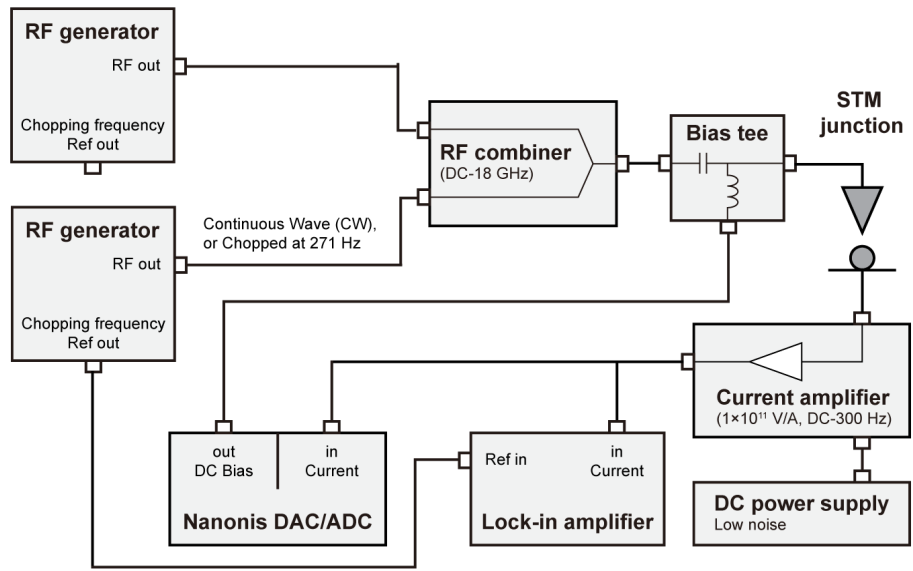

**Supplementary Fig. 9 | Schematic diagram of electric components and connection for all experiments.**